\documentstyle[12pt,aaspp4]{article}
\eqsecnum
\journalid{}{}
\articleid{}{}
\slugcomment{Accepted for publication in {\it The Astronomical Journal}}

\begin{document}

\title{Infrared Photometry and Dust Absorption in Highly Inclined Spiral Galaxies}

\author{Leslie E. Kuchinski\altaffilmark{1} 
and Donald M. Terndrup\altaffilmark{1,2}}
\affil{Department of Astronomy, The Ohio State University,\\
174 W. 18th Ave., Columbus, OH  43210 \\
Electronic mail: lek@payne.mps.ohio-state.edu,  terndrup@payne.mps.ohio-state.edu}

\altaffiltext{1}{Visiting Astronomer, Cerro Tololo Inter--American
Observatory, NOAO, which is operated by the Association of Universities
for Research in Astronomy, Inc.\ (AURA), under cooperative agreement
with the National Science Foundation.}
\altaffiltext{2}{Presidential Young Investigator.}

\begin{abstract}

We present $JHK$ surface photometry of 15 highly inclined, late-type
(Sab-Sc) spirals and investigate the quantitative effects of dust
extinction.  Using the $J - H$, $H - K$ two-color diagram, we compare
the color changes along the minor axis of each galaxy to the
predictions from different models of radiative transfer.  Models in
which scattering effects are significant and those with more than a
small fraction of the light sources located near the edge of the dust
distribution do not produce enough extinction to explain the observed
color gradients across disk absorption features. The optical depth in
dust near the plane as deduced from the color excess depends
sensitively on the adopted dust geometry, ranging from $\tau =  4$ to
15 in the visual band.   This suggests that a realistic model of the
dust distribution is required, even for infrared photometry, to correct
for dust extinction in the bulges of nearly edge-on systems.

\end{abstract}

\section{Introduction}

Observational studies of the bulges and inner disks of spiral galaxies 
are severely hampered by extinction due to dust in the disk.  Several 
recent works have demonstrated the importance of correctly treating 
extinction effects.
For example, while surface brightness profiles can be fit to
obtain morphological parameters, the presence of dust can
make the measurement of these parameters sensitive to wavelength
(\cite{byun94}; \cite{eva94}).  Surface photometry can also be used to
estimate the mass distribution of the inner galaxy in order to
understand dynamical influences on the bulge.  However, light blocked by
dust can exacerbate disagreements with kinematical mass estimates and
thus suggest the presence of more dark matter than is actually required 
to explain the rotation curve (\cite{bar94}).  The effects
of dust on broad-band colors are similar to those arising from changes
in metallicity and/or age, so the colors and color gradients provided
by multiwavelength images may contain little or no conclusive
information about stellar populations.  Clearly, an understanding of 
extinction effects is a necessary prerequisite to using the basic 
properties of bulges as indicators of their structure, stellar content, 
and history.

In late-type spirals, neither the quantity of dust nor its distribution
with respect to the stellar component is precisely determined.  Radiative
transfer models for plane-parallel and spherically symmetric geometries
predict that the amount of dust needed to produce a given
reddening depends sensitively on whether the dust lies in front of the
luminosity sources or is mixed with them in various ways
(\cite{jan94}; \cite{witt92}, hereafter WTC92).  Similar calculations 
for spiral galaxies consisting of a bulge, a disk, and a disk absorbing 
layer show that the observed extinction depends on both the optical 
depth in dust and the structural parameters of the absorption layer
(\cite{kod94}; \cite{oht95}).  These calculations and a recent model of the 
dust in NGC 4594 suggest that it may be necessary to represent the 
absorbing material with a ring structure (\cite{oht95}; \cite{ems95}). 
Scattering can also have a significant effect on the
observed surface photometry and colors, especially at wavelengths where
the albedo of dust grains is large (WTC92, \cite{byun94}, \cite{ems95}).
 Extinction effects
in models that include scattering tend to be less severe than in models
with no scattering because light scattered into the line of sight
partially compensates for absorption and reddening.  A suitable
representation of the dust component should be able to explain observed
reddening effects over a long spectral baseline.

We have begun an extensive study of the structure and kinematics, and 
evolutionary history  of bulges in highly inclined galaxies, some of
which show boxy or peanut-shaped morphology. Bulges in late-type spirals, 
which suffer from the most significant extinction effects, 
are of particular interest because the disk component dominates the
gravitational potential and therefore the history of the bulge and disk
 may be coupled. In this first paper, we present
infrared surface photometry and explore the characteristics of the dust
absorption features.  Our goal here is to provide quantitative 
estimates of the observed extinction and reddening in these systems.
Once we have determined how to correct for the effects of dust, we 
will investigate several proposed bulge formation mechanisms, including
disk instabilities, dissipational collapse from the protogalactic gas
cloud, or the accretion of small satellites (e.g., \cite{jon83}; 
\cite{qui86}; \cite{com90}; \cite{mer94}).  These processes 
would produce distinct signatures in bulge morphology and stellar 
populations, such as boxy shapes or the presence or absence of
metallicity gradients.  There are also
several clues which suggest that some bulges, especially boxy 
ones, are related to bars in the inner disks of the galaxies that harbor 
them.  Theoretical studies of resonant heating by bar instabilities
have concluded that one may see round, boxy, or peanut-shaped bulges
 depending on the angle between the bar and the observer (\cite{com90}).
Observational evidence for the link between boxy-peanut bulges 
and bars has been found in the line-of-sight velocity distributions for a
few peanut-shaped bulges, which show the expected kinematical signatures of a
bar (\cite{kui95}).  Additionally, our own
Galaxy has a boxy bulge (e.g., \cite{dwe95}) and shows a great deal of 
evidence for a bar in the inner disk, including studies of the spatial
distribution of tracer objects such as Miras (\cite{nak91};
\cite{wei92}; \cite{whi91}; \cite{whi92}; \cite{sta94}), models of the
asymmetric distribution of gas velocities in the $(\ell,v)$ plane
(\cite{bin91}), the asymmetry in the infrared surface brightness of the
bulge (\cite{bli91}; \cite{wei94}; \cite{dwe95}) and an excess of
microlensing events towards the galactic center compared to models
based on an axisymmetric bulge (\cite{pac94}).  
 We intend to use our galaxy surface photometry, corrected 
for extinction, to model the stellar mass distribution in the inner 
regions of highly inclined spirals in hopes of illuminating the connection
between bars and bulges.

This paper is organized as follows.  Sec. 2 describes the $JHK$ surface 
photometry observations, data reduction techniques, and the determination
of color profiles across the galaxy bulges.  In Sec. 3 we provide details of
the radiative transfer models with which our data are compared.  Sec. 4 
contains the quantitative results from a comparison of color changes along 
the minor axes of the bulges to model predictions.  In Sec. 5, we compare 
the color changes produced by dust to those implied by a change in stellar
populations.   A discussion of the results and some implications for 
the study of galaxian mass distributions are given in Sec. 6.

\section{Observations and Data Reduction}

\subsection{Sample Selection and Observing Procedures}

Our sample of 15 galaxies was selected specifically to study the boxy
bulge phenomenon in late-type spirals.  We restricted the sample to 
Hubble Types from Sab through Sc and considered inclinations $i > 65^\circ$
(the range in which the boxy bulge morphology is identifiable).
Eight galaxies in the sample have been identified in
the literature as having boxy or peanut-shaped bulges (\cite{jar86};
\cite{dd87}; \cite{sha90}).  The remainder were selected from the {\it
Third Reference Catalogue of Bright Galaxies}, hereafter the RC3
(\cite{dev91}), to span a similar range of absolute magnitudes,
major-axis dimensions, and inclination angles as the galaxies with
boxy-peanut bulges.  Table 1 lists some basic properties of galaxies in
our sample.

Near-infrared ($JHK$) images of the bulge and inner disk regions of
sample galaxies were obtained from 20--21 and 27--31 December 1993 with
the Ohio State Infrared Imager and Spectrograph (OSIRIS) on the 1.5 m
telescope at CTIO.  The details of the instrument are described by
\cite{dep93}.  OSIRIS has a $256 \times 256$ NICMOS 3 array detector
and was employed in its wide-field mode, which yielded a field of view
of 4.5\arcmin\ at a scale of $\approx 1.1$ arcsec pixel$^{-1}$.  All
observations were obtained under photometric conditions.

We observed each galaxy using an alternating sequence of sky and
on-source positions.  The sky and galaxy frames were taken with positional
dithering so that bad pixels in the array could be filtered out when
the frames were combined.  Exposure times were set to $\approx 4.9$ sec
in each filter, so that the count level of the sky plus galaxy nucleus
was always less than half the saturation level of 32,000 counts.  For
these exposure times, the typical sky levels in the $J$, $H$, and $K$
filters were 1000, 1500, and 5000 counts respectively. The total
on-source integration times for each galaxy in each filter are given in
Table 1.  The total integration time on the sky was approximately 2/3
of the total time on the source; sky observations were made at
intervals of 90 s during the galaxy observations.

\subsection{Data Reduction}

We performed all steps in the data reduction using the VISTA package
(\cite{sto88}).  The first step was to apply a linearity correction to
each raw data frame, which was a quadratic polynomial in pixel
intensity amounting to less than 2\% at the brightest typical exposure
levels.  All frames were then scaled to a common exposure time using
the true integration time\footnote{The length of sequential exposures
in OSIRIS varies by $\approx 0.1$ s.  OSIRIS records the exact exposure
time for each image, storing it as the value of a designated pixel.
The extraction of the exposure time was performed before the linearity
correction.} for each exposure.

We obtained multiple exposures of the nightly twilight sky in each
filter to serve as flatfield corrections. A zero-exposure bias frame
was subtracted from each twilight sky frame, then they were 
multiplicatively scaled to
match intensity levels and combined.  Initial tests of the flatfields
did not yield satisfactory results because there were small, but
significant, amounts of scattered light on one part of the detector.
We corrected for this scattered light using multiple observations of a
standard star on a grid of positions over the detector.  We then fit a
quadratic surface to the difference between the instrumental magnitude
of the star at each point on the grid and the average instrumental
magnitude measured on the unaffected parts of the detector.  The
r.m.s.  residual in the photometry after correction with the polynomial
was 3\%;  this was only slight larger than the scatter (2.5\%) measured
from observations of photometric standards (Sec. 2.3), suggesting that
the corrected flat fields were good to about 1\%.  Subsequent tests of
the constancy of the sky level on the galaxy frames indicated that
the accuracy of the flattening process was significantly better than
this (below).

We created average sky frames for each sequence of galaxy observations
from the sky exposures which immediately preceded and followed the
galaxy frames. These were combined using a median operation to remove
stellar images.  The appropriate average sky frame was subtracted from
each galaxy frame, which effectively removed not only the sky emission 
but also the dark current and some gross additive features caused by
scattered light.  The sky-subtracted frames were then divided by the
corrected twilight flat fields and multiplied by a bad pixel mask to
remove recurring bad pixels from subsequent analysis.  To retain the
correct count levels for error calculations, a constant equal to the
mean of the subtracted sky frame was added back onto each galaxy frame.
Note that the mean of the subtracted sky frame is not necessarily
identical to the sky value on the galaxy frame because sky levels
can fluctuate rapidly in the IR.  Determination of a sky value to 
subtract from the final processed galaxy image is discussed below.

All the frames for an individual galaxy were then aligned to a single
reference position using the centroids of 3 to 4 bright stars on the
frames.  No rotation or difference in scale between the $J$, $H$,
 and $K$ images
was evident in a comparison of stellar centroids, so only row and
column shifts were required.  The alignment was performed using
bilinear interpolation with masked bad pixels excluded from the
calculation.  Tests of the alignment routine using artificial stars
showed slight smoothing in the direction of the shift but no
significant change in the stellar FWHM.  The aligned galaxy frames were
additively scaled to match the sky levels, then combined using a
weighted average based on the photon statistics and detector
read noise.  At each pixel position, individual pixel values 
that were more than $5 \sigma $ from the mean
of all others in the stack were rejected from the calculation. 

The final reduction step before photometric calibration was to
determine and subtract an accurate sky value for each galaxy image.
None of the galaxies fill the frames and most have only a few bright
stars in the field.  Thus we could determine the overall sky level by
averaging the modal sky values taken in four different boxes on blank
regions of the frame.  The r.m.s. scatter in the sky levels in
different portions of the frame, which is a measure of the accuracy of
flat fielding and scattered-light removal, were $\approx 0.2\% $ of the
sky level at $J$ and $H$ (corresponding  21.1 and 20.7 mag arcsec$^{-2}$
respectively) and 0.1\% at $K$ (corresponding to 19.7 mag arcsec$^{-2}$).

\subsection{Photometric Calibration}

All of our calibrated magnitudes and colors are on the CTIO/CIT system.
Because nearly all of the faint \cite{eli82} CTIO/CIT standards were 
too bright to be observed with OSIRIS, we observed the faint UKIRT 
standards (\cite{cas92}.  We then converted these onto the CTIO/CIT 
system using the transformation of Casali \& Hawarden (1992).
 The UKIRT standards do not span a large enough range
in color to determine the color term between OSIRIS and the CTIO/CIT
system, so we determined a color term from observations of the IR 
standards of \cite{car94} and red stars 
in the Coalsack  obtained on 26 and 28 January 1994 at CTIO with the
same telescope and instrumental set-up (\cite {jon80}; \cite{ali95}).

Holding the color term (derived above) fixed, we obtained zero points,
airmass terms, and where appropriate, a UT term for each night of our
galaxy observations.  Although all nights were photometric, software
problems on the first two nights of the run resulted in an inaccurate
record of exposure times.  Snapshots of galaxies observed on these
two nights were taken later in the run and used for calibration.  
The airmass terms on the remaining five nights did not vary 
significantly and weather conditions were relatively constant, so we
used the same average airmass term on all nights.  The 
transformation equations for each night are of the form:
\begin{eqnarray}
K &=& k + 0.011 (J-K) + c_1 - 0.054 X \nonumber \\
J-K &=& 0.952 (j-k) + c_2 - 0.017 X [+ c*UT] \nonumber \\
H-K &=& 0.894 (h-k) + c_3 + 0.014 X \nonumber
\end{eqnarray}
where capital letters denote magnitudes on the CTIO/CIT system, small 
letters denote OSIRIS instrumental magnitudes, $c_1$, $c_2$, and $c_3$
are the zero points, and $X$ is the airmass.  The zero points remained 
constant within the errors from night to night, and
a UT term was required only on 28 December for the $J$ filter.  Figure
1 shows the residuals in $K$, $J - K$, and $H - K$ vs. color and
magnitude for all nights.  The r.m.s. residuals in the transformations
are $\leq 0.025$ mag, which we take to be the error in our calibration.

We compared photometry in synthetic apertures on our calibrated images
to published aperture photometry available in the literature for
several galaxies in our sample (\cite{aar82}; \cite{dev89};
\cite{aar89}; \cite{bot90}). Table 2 shows the differences in aperture
magnitudes in the sense ours {\it minus} published.  Our photometry
agrees with the previous values to within a few hundredths of a mag for
most of the apertures larger than 10\arcsec. Discrepancies in the
large aperture ($\geq 10\arcsec $) photometry for NGC 2613 and NGC 3717
are likely due to difficulties in correcting for the presence of
bright stars within the aperture.  For
apertures less that 10\arcsec\ in diameter, our photometry is
consistently brighter than the aperture values.  Photometry in small
apertures can be subject to significant seeing and alignment 
effects; \cite{ter94}
also note disagreement between synthetic aperture measurements from
arrays and single channel photometry for apertures $\leq 10\arcsec$.

\subsection{Color Maps and Color Profiles}

We computed color maps and profiles in $J - K$, $H - K$, and $J - H$
by aligning the appropriate pairs of images for each galaxy, then
computing the colors at each pixel.  Examples of the surface brightness
and color maps are shown in Figure 2, which displays maps in $K$ and 
$J-K$ for IC 2531.
Figure 3 displays $J - K$ color profiles along the minor axis for each
galaxy.  We set the minor axis position angle to $90^\circ$ from the
major axis position angle given in the RC3.  Colors were measured at
one-pixel intervals along a cut through the center of the galaxy, which
were measured in the $K$ band to minimize the effects of dust.) Because
the galaxy axes were not necessarily aligned with those of the
detector, each point along the minor axis was not typically centered on
a pixel.  The color at each location was therefore taken as the average
of surrounding pixels weighted by the distance of each pixel from the 
non-integer position at which the color is desired.
The error bars on the colors represent photon statistics in the
images, including the contribution from the night sky, but do not
include errors in the photometric calibration which (above) are about
0.025 mag. 

The color maps and profiles show red features in areas 
affected by absorption from dust in the disk, but the bulge 
colors in regions away from the dust lanes and galaxy nuclei 
remain relatively constant.
The most highly inclined systems in our sample -- NGC 1886, IC
2531, and NGC 3390 -- have red dust lanes along the major axis,
bisecting the bulge.  In systems at slightly lower inclinations, such
as NGC 1055, NGC 1589, and A0908--08, the red dust lane is displaced
from the major axis and one side of the bulge appears to be reddened by
disk dust.  Some of the galaxies, for example NGC 3717,
have very red nuclei.  The blue color near the center of NGC 1964 is
due to a foreground star superposed on the galaxy.  In general,
dust lanes in the edge-on galaxies appear as red peaks in
the profiles while systems of lower inclination show asymmetric bulge
colors with one side redder than the other.

As a function of inclination, our minor axis color profiles share many
of the qualitative characteristics of the $B - I$ profiles presented 
in Byun {\it et al.} (1994).  Those profiles are calculated from models 
of radiative transfer in systems consisting of a disk with a dust layer
and a dust-free bulge component.  We will not attempt to compare these
models to our IR data in a quantitative fashion because dust effects can 
be quite different at optical and infrared wavelengths.

\section{Models of Dust Absorption and Scattering}

In this section we describe several different models of radiative
transfer through dusty media that yield quantitative estimates of
reddening and extinction;  we compare the predictions of these models
to our color profiles in Sec. 4, below.  Total extinctions and color
excesses are predicted as a function of some measure of total dust
quantity, here the total $V$--band optical depth $\tau_{\rm V}$  
through the galaxy.  Models of
dust distributions can be divided into two types:  simple models with
analytical representations of radiative transfer and complex ones that
require a full numerical solution of the transfer equation.  Simple
models consider absorption but not scattering and can only be
constructed for a few specific dust geometries.  The complex models
used here, those of WTC92,  
treat both absorption and scattering for different dust geometries
representing various types of galaxies.  It is important to note that
all of the models discussed here have smooth distributions of dust,
whereas observations of the Milky Way and nearby galaxies show
irregular and patchy dust features.

\subsection{Analytic Models}

The simplest and most common representation of dust in galaxies is the
plane-parallel foreground screen model, in which an obscuring
layer of dust absorbs light from a source behind it.  In the absence
scattering, the total intensity observed through the dust is given by 
$I(\lambda) = I_0(\lambda) e^{-\tau_{\lambda}}$, where 
$I_0(\lambda)$ is the unattenuated intensity from the source and
 $\tau_{\lambda}$ is the optical depth.  The total extinction at
 any wavelength is then given by
\begin{eqnarray}
A_{\lambda} &\equiv& -2.5\log_{10}{I(\lambda) \over I_0(\lambda)}, \nonumber \\
            &=& 1.086 \tau_{\lambda}. \nonumber
\end{eqnarray}
Because the wavelength dependence of absorption by dust grains in other
galaxies is poorly known (but see \cite{jan94}), it is common to assume
a Galactic reddening law to express optical depth $\tau$ as a function
of wavelength.  Here we adopt the extinction law of
\cite{rie85}. Color excesses $E(\lambda_2 - \lambda_1) = A_{\lambda_2}
- A_{\lambda_1}$ can then be computed from the adopted reddening law.
Figure 4 shows the $K$--band extinction $A_{K}$ and
the color excesses $E(J - K)$ and $E(H - K)$ as a function of
$\tau_{\rm V}$ for the foreground screen model.  This predicts a large
degree of reddening from low optical depths because all of the light
must pass through the dust layer.

Another simple dust model is that of a uniform mixture of dust 
and stars.  \cite{wal88} derive:
\[
A_\lambda = -2.5 \log_{10}\left( {
         1 - e^{-\tau_{\lambda}} \over
         e^{-\tau_{\lambda}}
         } \right). \nonumber
\]
in the case of a slab with an equal mixture of stars and dust.  We plot
$A_{K}$, $E(J-K)$, and $E(H-K)$ as a function of $\tau_{\rm V}$
 for this model in 
Figure 4, where as above the Galactic extinction law was used to
express $\tau_{\lambda}$.  For the same optical depth, this model
produces less reddening than the foreground screen because light from
sources near the edge of the mixture escapes with little or no
extinction.

\subsection{Numerical Models}

WTC92 calculate numerical models of radiative transfer in spherically
symmetric systems and include the effects of both absorption and
scattering.  These models
are not ideal representations of spiral galaxies because of
their spherical shape and smooth dust distributions, but they
provide valuable insight into how different assumptions about the
distribution of the dust yield different color changes as
a function of total optical depth. The fraction of light
observed directly and the fraction scattered into the line of
sight are tabulated in WTC92 for several wavelengths ranging from
the ultraviolet to near-infrared. The extinction at some wavelength
$\lambda$ is given by the relation  
\begin{eqnarray}
A(\lambda) = -2.5\log_{10}\left( \frac {I_{direct}}{I_{0}} 
           + \frac {I_{scatt}}{I_{0}}\right) . 
\end{eqnarray}

We select three of the WTC92 dust geometries for use in our quantitative
comparison
of model and observed color excesses;  these are called the ``dusty galaxy,''
``starburst galaxy,'' and ``dusty galactic nucleus'' models.  The
dusty galaxy is a sphere of uniformly mixed stars and dust that differs
from the simple uniform mixture model due to the spherical geometry and
the inclusion of scattering.  Because light is scattered into the line 
of sight and because sources near the surface suffer little extinction,
there is much less reddening at each $\tau_{\rm V}$ than is produced
by the simple foreground screen.  (Recall that it is the extinction as 
a function of optical depth which distinguishes the various models.)
 The starburst galaxy has a centrally concentrated spherical
distribution of stars with a uniform sphere of dust embedded in it.
Most of the light in this model comes from the central dusty regions,
so reddening effects are more pronounced than for the dusty galaxy. 
The dusty galactic nucleus model contains a sphere of stars surrounded
by a spherical shell of dust, with no mixing of dust and light 
sources in either region.  This model differs from the simple
foreground screen mainly due to the inclusion of scattering, which
compensates for some of the
absorption to produce slightly less reddening than the screen model.
Extinctions and color excesses are calculated for $J$, $H$, and $K$ for
these models, and the results are presented as a function of $\tau_{\rm
V}$ in Figure 5.

\subsection{The Treatment of Scattering}

Although scattering is often neglected in favor of retaining simplicity
in dust models, it may have considerable effects on the observed
extinction characteristics.  Scattered light partially compensates for
absorption, and reddening effects on colors are less severe because
light of shorter (bluer) wavelengths is preferentially scattered into
the line of sight.  The role of scattering depends strongly on the
geometrical distributions of dust and stars as well as the dust grain 
albedo and angular scattering properties.  The models adopted here assume
albedos of 0.36, 0.28, and 0.20 for $J$, $H$, and $K$ respectively and 
phase function asymmetry parameters of 0.15, 0.04, and 0.00 in these 
bands.  Figure 6 shows the percentage
of observed light that is due to scattering into the line of sight
for several of the WTC92 models.  Contrary to what may be expected,
scattering effects are not largest at the highest optical depths
because so much light is absorbed in these cases that there is little
left to be scattered (WTC92).  This is evident from comparing
scattering at $\tau_{\rm V} = 1$ (upper panel) and $\tau_{\rm V} = 5$
(lower panel).  
We note that scattering is probably not significant for the infrared
photometry presented in this paper, since the scattering fraction for
$\tau_{\rm V} < 5$ is always less than 10\%.  However,  
the effects of scattering will be different at shorter wavelengths 
as extinction optical depths $\tau_{\lambda}$ are higher, scattering is 
less isotropic, and dust grain albedos may be higher (WTC92, but recent
models and observations (\cite{wit94}; \cite{kim94}) suggest a $K$ albedo
as high as the optical values of 60 to 70\% instead of the lower values
of 20 to 30\% in the standard dust mixture of \cite{dra84}).

Byun {\it et al.} (1994) find that scattering effects
can be ignored for galaxies with inclinations $i > 85^\circ$ 
because little light is scattered into the plane of the disk.
However, even at inclinations as high as $70^\circ$, neglect of 
scattering can contribute to errors of several $\times$ 0.1 mag in 
the predicted total $B$ band magnitude (\cite{byun94}).
On the other hand, Emsellem (1995) finds that scattering effects 
are necessary to simultaneously explain the $B$, $V$, $R$, and $I$ band 
attenuation in the dust lane of NGC 4594 ($i = 84^\circ$) and that much more
dust is required to explain the observed extinction if scattering is 
considered than if it is neglected.
In general, the above discussion suggests that estimates
 of the total extinction must be based on 
models with both absorption and scattering 
unless it is certain that a particular galaxy has a dust geometry 
or inclination for which scattering is unimportant.

\section{Comparison of Dust Models to the Minor-Axis Colors}

\subsection{Method of Analysis}

We now compare predictions from models of dust extinction to the
minor-axis colors
of our sample galaxies.  For each point along the minor axis which
experiences reddening within the galaxy, the intrinsic colors $(J -
H)_0$, $(H - K)_0$ would be observed at redder colors; for each dust
model the reddened minor-axis colors would be spread out along the path
given by $(J - H)_0 + E(J - H)$ and $(H - K)_0 + E(H - K)$ as a
function of $\tau_{\rm V}$.  In the simple foreground-screen model,
$E(J - H) \propto E(H - K)$ and so we have the traditional reddening
``vector'' in the two-color plot.  In other dust geometries or when
scattering is included, the ratio $E(J - H) / E(H - K)$ is not constant
and the behavior of the two colors with increasing optical depth
follows a curved path, which we will term the color ``trajectory'' of
each model.  Therefore the distribution of colors along the minor axis
in the two-color plane allows us to explore whether any of the simple
analytic or numerical models described above are appropriate models of
the effects of dust in our sample.

To compare the minor-axis colors to the dust models, we assume that the
extinction from the disk is negligible in the outer parts of each
bulge (away from the plane of the galaxy). Our assumption is
 supported by the color profiles in Figure 3, in
which the bulge colors are typically constant with minor-axis distance
away from the dust lane.  We use the color maps and minor-axis color 
profiles to select a location away from the plane in each galaxy which has
constant colors and small errors;  the colors at this point are taken to
be the intrinsic bulge colors $(J - H)_0$ and $(H - K)_0$.  We match each 
dust model at $\tau_{\rm V} = 0$ to these colors.  In galaxies
that are almost edge-on, for example IC 2531 and NGC 3390, the dust
lane bisects the bulge along the major axis and bulge colors are equal
within the errors on both sides.  For these systems we average the
outer bulge colors on either side of the dust lane.  Galaxies that are
not edge-on exhibit more reddening on the side obscured by the dusty
disk.  In this case, we adopt colors from the outer bulge regions on
the unobscured side.  Table 3 summarizes the adopted unreddened bulge
colors for each galaxy.

The outer-bulge colors in our sample compare favorably to the colors
measured for other galaxies.  Figure 7 shows the $(J - H)_0$, $(H -
K)_0$ values (filled points), along with the nuclear colors (from 
aperture photometry) for a
sample of Sc galaxies (open points) from Frogel (1985).  In this figure
only, our colors and Frogel's have been corrected for Galactic
reddening using the absorption-free polar cap model of \cite{san73} and
the \cite{rie85} extinction law.  That the colors for the two samples
are similar suggests that our assumption that the outer bulge colors
are relatively unreddened is not too far off.  Some of
the Sc nuclei are redder at both $J - H$ and $H - K$ than our
galaxies, possibly due to dust features contained in the apertures
(\cite{fro85}).

\subsection{Comparison to the Models}

Figure 8a--o show $J - H$, $H - K$ diagrams for galaxy colors along the
minor axis.  The colors are taken
from both sides of the galaxy center out to where the intrinsic color
was measured.  Each plot also shows the color trajectories for the
models discussed above.  For most galaxies, we assume that
 $\tau_{\rm V} = 0$ 
corresponds to the intrinsic colors measured in the previous 
section.  Exceptions to this assumption for some individual galaxies
are discussed below. 

Many galaxies in our sample, particularly the ones of lower inclination, 
do not have enough reddening to distinguish between different models 
because all of the models follow the same trajectory at low $\tau_{\rm V}$.
Three galaxies -- NGC 1325, NGC 2613, and NGC 2713 (Figure 8a--8c) -- 
have no identifiable dust lanes or have dust features located near
 the edge of the galaxy where the signal-to-noise is too low to 
determine reliable colors.
(For NGC 2613, in which the dust lane is visible at the outer edge 
of one side of the disk, we locate $\tau_{\rm V} = 0$ for the 
models at the blue edge of the dust feature and consider only the color 
change across these pixels.)
Four other systems -- NGC 1515, NGC 1589, NGC 1964, and
 IC 2469 (Figure 8d--8g) -- have dust features that are not
 red enough to extend into the regime 
on the color-color plot where different models can be distinguished.
As noted earlier, NGC 1964 has a star superposed near the center of the
galaxy that has bluer IR colors than the bulge itself.  Points bluer
than the adopted unreddened color of this galaxy (shown as asterisks in 
Figure 8f) are affected by the
presence of the star and are ignored in the model comparisons.  NGC
1589 has extremely red nuclear $H-K$ colors which are identifiable 
in the color-color plots as points with low errors (due to the high 
signal-to-noise in the center) lying far to the red of the model
predictions in $H-K$.  These points (asterisks in Figure 8e) are
ignored in the analysis.

The minor-axis colors of most of the remaining galaxies in our sample  -- 
NGC 1886, NGC 1888, ESO 489--29, A0908--08, IC 2531, and NGC 3390
(Figure 8h--8m) -- follow the trajectories described by
the simple foreground screen and uniform mixture models and the
starburst galaxy model of WTC92.  The large scatter of points for 
NGC 1886 also allows for the possibility that the dusty galaxy 
nucleus model could fit the data, but the trajectory of the data 
points does not follow the curve of this model.  NGC 3717 (Figure 8n),
like NGC 1589 above, has extremely red nuclear colors in $H-K$.
 Neglecting these points (asterisks in Figure 8n), 
we find that the reddening on the red side of the bulge (open points
in Figure 8n) is also described by the foreground
screen, uniform mixture, and starburst galaxy models.
We have located the $\tau_{\rm V} = 0$ point for the models at 
the bluest color of the red (obscured) side of the NGC 3717 
bulge because the colors on the red side follow a reddening 
trajectory while those on the blue side (solid points
in Figure 8n) are relatively constant.

The $JHK$ colors of NGC 1055 are quite different on the obscured 
and unobscured sides of the bulge.  Here we select two locations 
for $\tau_{\rm V} = 0 $ of the models, one for each side.  Clearly 
the bluest color of the red (obscured) side is not actually free of 
reddening, but this is not important because we are interested
in the change in color across the dust feature.
The curved trajectory of colors on the blue side of NGC 1055 
(open circles in Figure 8o) resembles the behavior of the
dusty galactic nucleus model at high $\tau_{\rm V}$, where it
strongly deviates from the other models by curving in the opposite 
direction.  As described in Sec. 3,
the dusty galactic nucleus model is basically a foreground screen
with the inclusion of scattering.  It is possible that in this case, 
we are observing disk stars through a
dust lane at an angle at which scattering effects are important.  The
outer-bulge colors we measured may not be completely free of reddening
even for the blue side of the bulge; this would explain why the
 model diverges from the
simple foreground screen at high $\tau_{\rm V}$ while our data diverge
at what we have called $\tau_{\rm V} = 0$ .  The colors on the red side of 
NGC 1055 (solid circles in Figure 8o) do not appear to follow any of 
the model trajectories.

Dust geometries with a significant fraction of light sources located
near the edges of the system never produce enough reddening to
explain the extremely red dust lane colors observed in
the most highly inclined galaxies in our sample.
 Because the bulk of the IR light seen in a highly inclined disk galaxy is
 located behind the dusty disk in the bulge, it is 
reasonable that models with
unobscured light sources do not match the observations as well as screen-type
models.
This is particularly apparent in the dusty galaxy model of WTC92:
the colors level off at some optical 
depth and there is little or
no more reddening with increasing $\tau_{\rm V}$.
The starburst galaxy model of WTC92 and the simple uniform model 
also show this saturation effect, but it occurs at redder colors than 
in the case of the dusty galaxy so these two models can still describe
many of the observed dust lane colors.

For two cases of models with analogous dust distributions, we find
that the simple models with no scattering 
provide a better description of the color trajectory than the numerical 
models that include scattering.  The simple uniform model appears to be 
a much better fit to the observations than the analogous dusty galaxy 
model of WTC92, which has a trajectory that curves sharply blueward in 
$J-H$ at high $\tau_{\rm V}$. Likewise, the simple foreground screen model
lies along the observed color trajectories while the analogous dusty 
galactic nucleus model of WTC92 has a color trajectory that curves 
away from the observations and the other models considered here.
These results are not surprising:  scattering is not expected to 
be important in the IR for the highly inclined galaxies in our sample
because most of the light is from red bulge stars located behind the 
``screen'' of the dusty disk and thus little is scattered towards the 
observer.  However, we note that the mixture of dust and light sources
that would be seen in a face-on system or in an edge-on galaxy at blue
wavelengths (where the young stars in the spiral arms are visible)
would produce more scattering and would also require a different model
geometry.

\subsection{Dust Optical Depth in Highly Inclined Galaxies}

We now measure maximum color excesses in galaxy dust lanes to estimate the
range of total $V$ band optical depth predicted by various dust
models.  The sample is limited to those galaxies with clearly defined
dust lanes and $JHK$ colors that follow the model reddening
trajectories from the intrinsic colors out to the reddest values.
Columns 3--5 of Table 4 give the maximum $E(J - K)$, $E(H - K)$ and
$E(J - H)$ measured along the minor axis.  Corresponding $V$ band optical
depths at each color for each appropriate dust model are given in
Columns 6--8 of Table 4.  (For NGC 1589, we ignore the extremely red
nuclear points discussed above and use the reddest point in the dust
lane to calculate color excesses.) The values of $\tau_{\rm V}$
predicted from the three different colors of a single galaxy are
similar, as is expected if the model adequately describes the observed
colors.  We take an average of these three values as the estimated
$\tau_{\rm V}$ for the galaxy and present this and the corresponding
$K$ band extinction A$_{\rm K}$ for each case in columns 9--10 of
Table 4.

It is clear from Table 4 that the total optical depth in dust needed to
produce the observed reddening is highly sensitive to geometry  (the
models all have the same dust properties, and thus differ primarily in
geometry).  The simple foreground screen model provides a {\it lower} limit
on the optical depth; other geometries require more dust to produce the
same reddening because sources near the edge of the system do not
interact with much dust and/or scattering partially compensates for
absorption.  For the highly inclined galaxies in Table 4, the predicted
$\tau_{\rm V}$ from the screen model ranges from 1.6 to 2.7, while
corresponding $\tau_{\rm V}$ from the uniform mixture and starburst
galaxy models are much higher:  3.7 to 7.5.  For two of the edge-on
galaxies, NGC 1886 and IC 2531, only the screen model produces enough
 reddening and the predicted
values of $\tau_{\rm V}$ are $\approx 4.5$. The colors of NGC 3390 can
be represented by either the screen or uniform mixture models:  the
screen model predicts $\tau_{\rm V}$ $\approx 4.5$ and the uniform
mixture yields $\tau_{\rm V}$ $\approx 15$.  All of these values imply
that the dusty regions of highly inclined galaxies are optically thick
at visible wavelengths, but whether or not they would be optically thin
at face-on inclinations depends on which model is used to predict
$\tau_{\rm V}$ and the assumed ratio of dust scale height to disk scale 
length.

Figure 9 shows the maximum measured dust lane color excess as a function
of distance from the galaxy center along the {\it major} axis.
These values are determined from color profiles of cuts 
parallel to the minor axis: $E(J-K)_{\rm max}$  is measured for each 
profile just as it was for the minor axis above.  Because the 
signal-to-noise ratio is usually too low to measure an intrinsic 
color for cuts parallel to the minor axis,
the intrinsic bulge colors derived in Sec. 4.1 are used.  (Although
many of these cuts pass primarily through the disk rather than the
bulge, we note that our color maps show similar $J-K$ for bulge and
disk.  \cite{ter94} also find bulges and disks to have similar $J-K$
colors.)  The three edge-on galaxies of Table 4 show a strong peak in central
reddening; this could be due to a concentration of dust at the galaxy
center and/or an extremely red nuclear stellar population.  In regions
away from these three galaxies' centers and in the other three galaxies shown
in Figure 9, the maximum dust lane reddening remains relatively
constant.  In some cases there appears to be a slight tendency 
for $E(J-K)_{\rm max}$ to  diminish with
increasing galactocentric distance, which could be due to the fall-off
of an exponential disk of dust. Using A$_{\rm V,max}$ instead of 
$E(J-K)_{\it max}$, Jansen {\it et al.} (1994) 
find a rapid drop in the maximum dust lane extinction at
 increasing distance from the center for several highly inclined galaxies. 
 If such a decrease in optical depth does occur in our galaxy
sample, it may be noticeable at greater distances from the center or in
the more dust-sensitive optical colors.

\section{Dust and Stellar Population Studies}

Figure 10 shows the changes in $JHK$ colors produced by age and
metallicity, from the models of \cite{wor92}, along with the reddening
trajectories of the three models that best describe galaxy dust
lane colors.  It is clear that for these colors, the reddening 
produced by the dust models lies parallel to the color changes that 
would result from gradients in metallicity and age.    
Because the quantity of dust is not well constrained, it
is impossible to determine how much of a color change along the
observed trajectory is due to dust and how much is due to a change
in the stellar population.  This unfortunate situation greatly limits
the use of broad-band photometry for stellar population studies.

Once a dust model is selected to describe reddening within a galaxy, 
it may be possible to find specific color-color diagrams in which 
the effects of extinction can be distinguished from changes in
stellar population (e.g. \cite{wit95}).  However, the difficulty of
using colors as population indicators is also exacerbated by degeneracies
in the changes produced by metallicity and by age.  Spectral line
studies, rather than colors, appear to be a more effective tool to
disentangle the effects of dust, metallicity, and age and to study
population gradients in spiral bulges.

\section{Summary and Discussion}

We compare $JHK$ colors across the dust lanes of highly inclined late
type spirals to predictions from several different dust models.  All of
the representations available for quantitative comparisons have smooth
dust distributions (no patchy features) and none has precisely the
geometry of spiral galaxies.  Two common analytical models, the
foreground screen and the uniform mixture, provide good descriptions of
the observed reddening.
The starburst galaxy model of WTC92 also adequately describes the
observed color profiles.  Dust geometries with a significant fraction
of unobscured light sources such as the WTC92 dusty galaxy model never
produce enough reddening to match the observed colors in the reddest disk
absorption features seen in our galaxy sample.
Although the dusty galactic nucleus and dusty galaxy models of
WTC92 differ from the simple analytical screen and uniform models
only in the inclusion of scattering and the spherical (rather than
plane parallel) geometry, they predict very different 
$JHK$ colors as $\tau_{\rm V}$ increases.  The lack of agreement 
between the models with scattering and the IR dust lane colors of 
highly inclined galaxies suggests that in the IR, scattering
effects are not significant for the objects that make up our sample.
However, as noted in Sec. 4.2, both scattering and models that 
include some unobscured sources may be required to describe face-on
systems at any wavelength or galaxies of any inclination at 
blue wavelengths where young disk stars are visible.
Such models would be particularly important for very late type spirals
in which a small bulge may no longer be the dominant light source.

Because three different dust models each provide a satisfactory
description of the galaxy colors, we cannot derive a single value for
the total optical depth of dust.  Instead, we use the three results to
place limits on this quantity. In all cases, the simple foreground
screen model that is traditionally used to discuss dust extinction
provides the lower limit on optical depth $\tau_{\rm V}$ because the
light sources interact with all of the dust and there is no scattering
to compensate for absorption.  Our predicted $\tau_{\rm V}$ for edge-on
galaxies ranges from $\approx 4.5$ for the screen model up to 15 for
the uniform mixture.  Values for dust lane optical depth in galaxies
that are not exactly edge-on range from near 2 for the screen model to
4 to 7.5 for the uniform mixture and starburst galaxy models.  These
limits provide guidelines for understanding galaxy optical depths, but
they can be narrowed to more precise values only with a better
understanding of the geometry and scattering properties of the dust.
That models with quite different $\tau_{\rm V}$ give the same reddening
leads us to conclude that an estimate of the absorption-free
light distribution in {\it any} edge-on galaxy {\it requires a proper
treatment of radiative transfer.}  In our future papers, we will
analyze our optical data for these galaxies using the procedures developed
in this paper.  This will help determine which dust models can describe
the observed colors over a long spectral baseline, including the more
dust-sensitive blue wavelengths.

There are inherent difficulties in our method of selecting intrinsic
bulge colors that should not be overlooked, but these have little
impact on our qualitative conclusions about dust models.  By assuming
that the outer bulge colors are not reddened, we imply that there is
little or no dust mixed in with the bulge component.  The constant
outer bulge colors evident in our data and the similarity between our
intrinsic colors and the colors of lower inclination Sc galaxies
from \cite{fro85} suggest that there is indeed little or no reddening
in the outer bulge.  We have also assumed that the intrinsic colors of
the inner and outer bulge are the same in order to measure a color
excess in the dust lane.  This neglects any change in color due to a
stellar population gradient.  If we assume a metallicity change of
$\approx 0.4$ dex, consistent with that measured for the Milky Way
bulge (e.g., Terndrup 1988; Frogel {\it et al.} 1990; Terndrup
{\it et al.} 1990; Tiede {\it et al.} 1995), the
corresponding color change would be $\approx 0.15$ in $J-K$ based on the
stellar population models of \cite{wor92}.  This is much smaller than
the total color change across the dust lanes (see Table 4), suggesting
that the observed $J-H$ and $H-K$ color changes are for the most part
due to dust extinction.  Finally, it is possible that blue light
scattered from the side of the disk located behind the bulge may
contribute to a slight bluing of the unobscured side.  Although all of
the factors mentioned here may contribute to errors in the total
$\tau_{\rm V}$ values we measure, they do not affect the basic
trajectories in the color-color diagrams and thus will not change our
conclusions about which models best describe the observations.

Our study of dust extinction in late-type spirals can help pinpoint
where mass models based on the light distribution are likely to be
inaccurate, an important step in understanding dynamical influences on 
the bulge and inner disk. 
We predict values of A$_{\rm K} \approx 0.5 - 0.6$ at the
centers of edge-on galaxies and $0.2-0.3$ in the dust lanes of highly
inclined galaxies that are not exactly edge-on (Table 4).  In fact,
once a plausible dust model is specified, the extinction A$_{\lambda}$
at any wavelength can be determined for any location based on the
reddening at that point.  These extinction values can be used to
understand how the mass to light ratio ($M/L$) changes over galaxy dust
features or even to crudely correct the light distribution so that a
constant $M/L$ can be invoked.  Combined with kinematical data, an
improved picture of the stellar light distribution in the inner regions
of spirals will shed light on the interaction between disks and bulges.
It will then be possible to address questions about the history of the 
bulge and disk components, the relation between bars and bulges, and 
why some galaxies have boxy or peanut-shaped bulges.

\acknowledgments

We wish to thank Darren DePoy, Jay Frogel, Andy Gould, Richard Pogge, Alice
Quillen, and Kristen Sellgren for advice and helpful discussions.
We are also grateful to Jay Frogel for providing Sc galaxy colors
in machine-readable form.  Babar Ali, Darren DePoy, Alice Quillen,
and Jay Elias provided
observations of standard stars for photometric calibration.  Our thanks
go also to the staff of the Cerro Tololo Inter-American Observatory,
who provided able assistance during the observations.
Last but certainly not least, we thank the anonymous referee for several
helpful comments.  This project was
supported in part by grant AST91-57038 from the National Science
Foundation.  OSIRIS is funded by grants AST90-16112 and AST92-18449 
from the National Science Foundation.  Galaxy research at OSU is funded 
by grant AST92-17716 from the National Science Foundation.

\clearpage

\begin{figure}
\caption{Magnitude and color residuals for the photometric
calibrations, in the sense published values for the standard 
stars {\it minus} our values.  Photometry is on the CTIO/CIT
system.  Data are shown for all 5 nights on which calibrations
were carried out.}
\end{figure}

\begin{figure}
\caption{Contour plots of the $K$-band surface photometry and 
$J-K$ color map for IC 2531, an example of an edge-on galaxy 
in our sample.  North is on the bottom and East is to the left.}
\end{figure}

\begin{figure}
\caption{Color profiles along the minor axis of each galaxy.  The
errorbars represent uncertainties due to photon statistics and sky
noise in the $J$, $H$, and $K$ images; they do not include calibration
errors.  Galaxy inclination angles estimated from Bottinelli 
{\it et al.} (1983) are given in the upper right of each panel.}
\end{figure}

\begin{figure}
\caption{$K$ band extinction A$_{\rm K}$ and color excesses $E(J-K)$
and $E(H-K)$ as a
function of $\tau_{\rm V}$ for the simple foreground screen and uniform
mixture models.  The Galactic reddening law of Rieke \& Lebofsky
(1985) has been used to express $\tau(\lambda) $ as a function of
$\tau_{\rm V}$ for the model calculations.}
\end{figure}

\begin{figure}
\caption{$K$ band extinction A$_{\rm K}$ and color excesses $E(J-K)$
and $E(H-K)$  as a
function of $\tau_{\rm V}$ for the radiative transfer models of WTC92.
The scale identical to that of Fig. 4 to facilitate comparison.}
\end{figure}

\begin{figure}
\caption{Fraction of observed light that has been scattered into the
line of sight as a function of wavelength for WTC92 dust
models with $\tau_{\rm V}$ = 1 (upper panel) and $\tau_{\rm V}$ = 5
(lower panel).}
\end{figure}

\begin{figure}
\caption{Adopted outer-bulge $J-H$ and $H-K$ colors of galaxies in
 our sample (filled squares) and nuclear colors of Sc galaxies 
from Frogel (1985) (open circles).  The
locations of dwarf stars and giants in globular clusters, the local
field, and Baade's Window are shown for comparison.}
\end{figure}

\begin{figure}
\caption{$J-H$, $H-K$ diagrams for galaxy minor axis colors.  Errors in
the colors are calculated from photon statistics and sky noise on the
$J$ $H$, and $K$ images.  Typical calibration errors are shown in the
lower right corner of each plot.  The reddening trajectories predicted
by different dust models are shown as lines originating at the
$\tau_{\rm V} = 0$ locations discussed in Sec. 4.2.
For NGC 1055 and NGC 3717, which have different color trajectories on 
the obscured and unobscured sides of the bulge, the colors on one side 
are denoted by open points and those on the other side are denoted by 
closed points.  Central points of NGC 1589, NGC 1964, and NGC 3717 are
ignored as described in the text; these are denoted with starred points
on the plots.}
\end{figure}

\begin{figure}
\caption {Maximum $J-K$ color excess in cuts across the dust lane 
as a function of distance of that cut from the galaxy center (along the
major axis).  Errorbars are calculated from the errors in the intrinsic
and reddest colors, which reflect photon statistics and sky noise on
the $J$ and $K$ images.}
\end{figure}

\begin{figure}
\caption {Changes in IR colors produced by age (left panel) and
metallicity (right panel) from the stellar population models of Worthey
(1992).  Ages range from 8 to 17 Gyr for [Fe/H] = 0.0, metallicity is
from [Fe/H] = $-$1.0 to +0.25 for an age of 12 Gyr.  Reddening
trajectories predicted by the dust models that describe highly inclined
spirals are shown on both plots, with $\tau_{\rm V} = 0$ located at the
bluest (youngest or most metal poor) stellar population colors.}
\end{figure}

\clearpage
\begin{planotable}{llccccccc}
\tablenum{1}
\tablewidth{0pt}
\tablecaption{Basic Galaxy Data and Log of Observations}
\tablehead{
\colhead{Galaxy} & \colhead{Type\tablenotemark{a}} & \colhead{B/P\tablenotemark{b}} 
& \colhead{$B_T$\tablenotemark{a}} & \colhead{incl.\tablenotemark{c}}
& $\log D_{25}$\tablenotemark{a,d}& \colhead{$J$\tablenotemark{e}} & \colhead {$H$\tablenotemark{e}} & \colhead{$K$\tablenotemark{e}}
}
\startdata
NGC 1055      & SBb  & yes & 11.40 & 72 & 1.88 & 585 & 585 &  700 \nl
NGC 1325      & Sbc  & no  & 12.22 & 73 & 1.67 & 585 & 585 &  700 \nl
NGC 1515      & Sbc  & no  & 12.05 & 82 & 1.72 & 585 & 585 &  700 \nl
NGC 1589      & Sab  & yes & 12.80 & 75 & 1.50 & 585 & 585 &  700 \nl
NGC 1886      & Sbc  & yes & 13.60 & 90 & 1.49 & 390 & 390 &  700 \nl
NGC 1888      & SBc  & yes & 12.83 & 77 & 1.48 & 585 & 585 &  700 \nl
NGC 1964      & Sb   & no  & 11.58 & 70 & 1.75 & 585 & 565 &  700 \nl
ESO 489$-$29  & Sbc  & yes & 13.33 & 90 & 1.47 & 585 & 585 &  820 \nl
NGC 2613      & Sb   & no  & 11.16 & 80 & 1.86 & 315 & 315 &  700 \nl
NGC 2713      & SBab & no  & 12.72 & 68 & 1.56 & 585 & 585 &  700 \nl
A0908$-$08    & Sb   & no  & 11.91 & 81 & 1.63 & 585 & 585 &  700 \nl
IC 2469       & SBab & yes &\nodata& 85 & 1.67 & 390 & 315 &  700 \nl
IC 2531       & Sc   & yes & 12.90 & 90 & 1.84 & 585 & 585 &  700 \nl
NGC 3390\tablenotemark{f}      & Sb   & yes & 12.85 & 90 & 1.55 & 700 & 700 & 1050 \nl
NGC 3717      & Sb   & no  & 12.24 & 88 & 1.78 & 585 & 585 &  820 \nl
\enddata
\tablenotetext{a}{Data from the RC3.}
\tablenotetext{b}{B/P = boxy or peanut-shaped bulge identified in Jarvis 
(1986), de Souza \& dos Anjos (1987), or Shaw {\it et al.} (1990).}
\tablenotetext{c}{Inclinations estimated from Equations 1 and 2 in Bottinelli {\it et al.} (1983).}
\tablenotetext{d}{log $D_{25}$ in units of 0.1 arcmin.}
\tablenotetext{e}{$J$, $H$, and $K$ exposure times in seconds.}
\tablenotetext{f}{NGC 3390 is classified as either S0$_3$ (S0 with dust lane) 
or Sb in Sandage \& Bedke (1994).} 
\end{planotable}

\clearpage
\begin{planotable}{lrccc}
\tablenum{2}
\tablewidth{0pt}
\tablecaption{Comparison to Published Aperture Photometry}
\tablehead{
\colhead{Galaxy} & \colhead{Aperture} & 
\colhead{$\Delta J$\tablenotemark{a}} & \colhead {$\Delta H$\tablenotemark{a}} & \colhead{$\Delta K$\tablenotemark{a}} 
}
\startdata
NGC 1055\tablenotemark{b}  &  81.2 & \nodata & $-0.05$ & \nodata \nl
          & 110.6 & \nodata & $-0.01$ & \nodata \nl
NGC 1325\tablenotemark{b} &  55.8 & \nodata & $-0.02$ & \nodata \nl
          &  70.1 & \nodata & $-0.01$ & \nodata \nl
NGC 1964\tablenotemark{b,c}  &   3.6 & \nodata & $-0.05$ & $-0.14$ \nl
          &   5.3 & \nodata & \nodata & $-0.18$ \nl
          &   7.2 & \nodata & \nodata & $-0.25$ \nl
          &   9.3 & \nodata & $-0.22$ & $-0.29$ \nl
          &  53.4 & \nodata & $-0.03$ & \nodata \nl
          &  81.2 & \nodata & $-0.10$ & \nodata \nl
          & 105.0 & \nodata & $-0.05$ & \nodata \nl
          & 110.6 & \nodata & $-0.10$ & \nodata \nl
NGC 2613\tablenotemark{b}  &  83.6 & \nodata & $-0.17$ & \nodata \nl
          & 110.0 & \nodata & $-0.19$ & \nodata \nl
IC 2531\tablenotemark{d}   &  59.2 & \nodata & $-0.07$ & \nodata \nl
          &  69.5 & \nodata & $-0.09$ & \nodata \nl
NGC 3390\tablenotemark{e}  &  10.0 & $-0.18$ & $-0.19$ & $-0.20$ \nl
NGC 3717\tablenotemark{b}  &  50.4 & \nodata & $-0.15$ & \nodata \nl
          &  81.2 & \nodata & $-0.15$ & \nodata \nl
\enddata
\tablenotetext{a}{$\Delta $mag in the sense (ours--published).}
\tablenotetext{b}{Aaronson {\it et al.} 1982.}
\tablenotetext{c}{Devereux 1989.}
\tablenotetext{d}{Aaronson {\it et al.} 1989.}
\tablenotetext{e}{Bothun \& Gregg 1990.}
\end{planotable}

\begin{planotable}{lrrrrrrr}
\tablenum{3}
\tablewidth{0pt}
\tablecaption{Adopted Intrinsic Bulge Colors}
\tablehead{
\colhead{Galaxy} & \colhead{$r(i)$\tablenotemark{a}} & 
\colhead{$J - K$} & \colhead{error} &
\colhead{$H - K$} & \colhead{error} &
\colhead{$J - H$} & \colhead{error}
}
\startdata
NGC 1055      & $-6.6$    & 0.91 & 0.04 & 0.30 & 0.04 & 0.61 & 0.03 \nl
NGC 1325      & $+8.8$    & 0.81 & 0.17 & 0.16 & 0.15 & 0.66 & 0.12 \nl
NGC 1515      & $+5.5$    & 0.97 & 0.04 & 0.27 & 0.03 & 0.71 & 0.03 \nl
NGC 1589      & $+8.8$    & 0.92 & 0.08 & 0.22 & 0.06 & 0.70 & 0.06 \nl
NGC 1886      & $-4.4$    & 0.82 & 0.11 & 0.21 & 0.07 & 0.61 & 0.10 \nl
NGC 1888      & $+5.5$    & 0.98 & 0.05 & 0.27 & 0.04 & 0.71 & 0.04 \nl
NGC 1964      & $-6.6$    & 0.98 & 0.03 & 0.28 & 0.02 & 0.70 & 0.02 \nl
ESO 489$-$29  & $-5.5$    & 0.89 & 0.07 & 0.24 & 0.06 & 0.65 & 0.06 \nl
NGC 2613      & $+7.7$    & 0.92 & 0.11 & 0.21 & 0.05 & 0.71 & 0.12 \nl
NGC 2713      & $-7.7$    & 1.00 & 0.05 & 0.26 & 0.04 & 0.74 & 0.04 \nl
A0908$-$08    & $+8.8$    & 0.88 & 0.06 & 0.23 & 0.05 & 0.66 & 0.04 \nl
IC 2469       & $-4.4$    & 0.95 & 0.05 & 0.27 & 0.02 & 0.68 & 0.05 \nl
IC 2531       & $\pm 6.6$\tablenotemark{b} & 0.96 & 0.10 & 0.27 & 0.09 & 0.70 & 0.08 \nl
NGC 3390      & $\pm 6.6$\tablenotemark{b} & 0.92 & 0.06 & 0.25 & 0.05 & 0.67 & 0.05 \nl
NGC 3717      & $-8.8$    & 0.85 & 0.04 & 0.20 & 0.03 & 0.65 & 0.03 \nl
\tablenotetext{a}{$r(i)$ = distance in arcsec from center (along minor axis) at which
intrinsic colors are measured.}
\tablenotetext{b}{For edge-on galaxies, colors on both sides of the bulge are 
averaged.}
\enddata
\end{planotable}

\clearpage
\begin{planotable}{llrrrrrrrr}
\tablenum{4}
\tablewidth{0pt}
\tablecaption{Total Optical Depth in Galaxy Dust Lanes}
\tablehead{\colhead{Galaxy} & \colhead{Model} & \colhead{$E_{\rm JK}$\tablenotemark{a}}
 & \colhead{$E_{\rm HK}$\tablenotemark{a}} & \colhead{$E_{\rm JH}$\tablenotemark{a}} & \colhead{$\tau_{\rm V,JK}$\tablenotemark{b}}
 & \colhead{$\tau_{\rm V,HK}$\tablenotemark{b}} & \colhead{$\tau_{\rm V,JH}$\tablenotemark{b}} & 
\colhead {$\langle \tau _{\rm V}\rangle $\tablenotemark{c}} & \colhead{$\langle A_{\rm K} \rangle $\tablenotemark{c}}
}
\startdata
 & & & & & & & & & \nl
Edge-On &  &  &  &  &  &  &  &  &  \nl
Galaxies &  &  &  &  &  &  &  &  &  \nl
 & & & & & & & & & \nl
NGC 1886 & screen &  0.80 & 0.31 & 0.50 & 4.36 & 4.49 & 4.29 & 4.38 & 0.53 \nl 
 & & & & & & & & & \nl
IC 2531 & screen & 0.81 & 0.30 & 0.52 & 4.40 & 4.33 & 4.44 & 4.39 & 0.53 \nl
 & & & & & & & & & \nl
NGC 3390 & screen & 0.79 & 0.35 & 0.44 & 4.29 & 5.07 & 3.82 & 4.39 & 0.54 \nl
NGC 3390 & uniform & 0.79 & 0.35 & 0.44 & 14.89 & 15.37 & 14.24 & 14.84 & 0.78 \nl
 & & & & & & & & & \nl
Highly &  &  &  &  &  &  &  &  &  \nl
Inclined &  &  &  &  &  &  &  &  &  \nl
Galaxies &  &  &  &  &  &  &  &  &  \nl
 & & & & & & & & & \nl
NGC 1589 & screen & 0.31 & 0.11 & 0.20 & 1.65 & 1.55 & 1.71 & 1.64 & 0.20 \nl
NGC 1589 & uniform & 0.31 & 0.11 & 0.20 & 3.77 & 3.37 & 4.04 & 3.73 & 0.22 \nl
NGC 1589 & starburst & 0.31 & 0.11 & 0.20 & 3.98 & 3.57 & 4.20 & 3.92 & 0.24 \nl
 & & & & & & & & & \nl
ESO 489-29 & screen & 0.43 & 0.15 & 0.28 & 2.31 & 2.18 & 2.38 & 2.29 & 0.28 \nl
ESO 489-29 & uniform & 0.43 & 0.15 & 0.28 & 5.63 & 4.94 & 6.17 & 5.58 & 0.32 \nl
ESO 489-29 & starburst & 0.43 & 0.15 & 0.28 & 6.02 & 4.97 & 6.91 & 5.97 & 0.36 \nl
 & & & & & & & & & \nl
A 0908-08 & screen & 0.49 & 0.18 & 0.31 & 2.66 & 2.66 & 2.66 & 2.66 & 0.32 \nl
A 0908-08 & uniform & 0.49 & 0.18 & 0.31 & 6.78 & 6.24 & 7.21 & 6.75 & 0.38 \nl
A 0908-08 & starburst & 0.49 & 0.18 & 0.31 & 7.33 & 6.41 & 8.18 & 7.31 & 0.43 \nl
\enddata
\tablenotetext{a}{$E_{\rm JK} = E(J-K)$, $E_{\rm HK} = E(H-K)$, $E_{\rm JH} = E(J-H)$.}
\tablenotetext{b}{$\tau_{\rm V,JK} = \tau_{\rm V}$ determined from $E(J-K)$, likewise for other colors.}
\tablenotetext{c}{Values for $\langle \tau_{\rm V}\rangle $ and $\langle A_{\rm K}\rangle $ are averages of the values determined for each of the 3 colors.}
\end{planotable}

\end{document}